# Electrical Conductivity of Thin-Film Composites Containing Silver Nanoparticles Embedded in a Dielectric Teflon® AF Matrix

*Haoyan Wei, Hergen Eilers*[*]

Applied Sciences Laboratory, Institute for Shock Physics, Washington State University, Spokane, WA 99210, USA

**ABSTRACT**

Thin-film nanocomposites, consisting of silver nanoparticles embedded in a dielectric Teflon® AF matrix, were synthesized using vapor phase co-deposition. The electrical conductivity of these composites was measured in-situ as a function of film thickness at various metal concentrations. At low metal concentrations (<30%), dielectric behavior and very little change with film thickness was observed. At moderate to high silver loadings (30-80%) a large increase in electrical conductivity was observed as the films grew thicker. As the thickness increased further, the conductivity flattened out. At very high silver content (>90%), fragmented fractal nanoclusters were able to further interconnect to achieve the percolation process and eventually evolve into a metallic continuum with dielectric polymer inclusions.

## 1. Introduction

Composites consisting of polymers with embedded nanostructured metallic particles have unique properties and are of particular interest for optical, electrical, and opto-electronic applications [1-4]. These composite materials exhibit unique optical characteristics originating from the strong interaction between incident light and metallic nanoparticles. This interaction results in collective oscillations of electron clouds, so called surface plasmons, at the interface of the metallic nanoparticles and the dielectric matrix. The resonance frequency of this interaction is strongly dependent on the metal, the surrounding dielectric medium, as well as the size and shape distribution of the nanoparticles.

With increasing metal loading, initially isolated metallic nanoparticles start to coalesce, leading to the formation of nanoclusters with irregular shapes and a broader size distribution. Such fractal structures can greatly extend the light absorption from the ultraviolet (UV) and visible (VIS) wavelength region into the infrared (IR) wavelength region [5]. The tunability of the optical properties of these metal-dielectric composites makes them suitable for a variety of applications including surface-enhanced Raman scattering (SERS) sensors [6], color filters [7-9], and all optical switching [10].

The surface plasmons can lead to intense electromagnetic fields in the dielectric between metallic nanoparticles. This effect has been applied in photovoltaic cells [11, 12] and light-emitting diodes (LEDs) [13, 14], enhancing the device efficiency due to strong coupling of the light to the interface (absorption or emission region, respectively). More recently, direct employment of metallic nanoparticles as photosensitizers was demonstrated [11, 15]. It was shown that excited surface plasmons can eject electrons into a surrounding conductive medium resulting in effective charge separation.

We recently presented significant progress in tailoring the optical properties of Ag/Teflon® AF nanocomposites by varying the concentration, and the size and shape distribution of embedded silver nanoparticles. At silver concentrations near the percolation threshold, silver nanoclusters with all sizes and shapes form. Such fractal networks resulted in an unusually large broadband visible to infrared absorption range (400 nm – 2500 nm and above) [4], which is of particular interest for multi-spectral sensor applications. We also demonstrated that the absorption profile of Ag/Teflon® AF composites can be tailored to closely match the solar radiation spectrum [16], indicating that these composites have potential use as next-generation photovoltaic cells.

The synthesis of such nanocomposites typically involves the deposition of metallic nanoparticles into a dielectric matrix. Polymeric matrices are of particular interest due to their relatively low cost and easy processability. The electrical properties of such composites are closely related to the morphology of the embedded metallic nanostructures, which are dependent upon both film thickness and metal concentration.

Previous studies of electrical properties of metal-filled polymers focused on the effect of metal concentration in films with thicknesses in the micrometer or larger range.

We report here on the correlation between the electrical conductivity of Ag/Teflon® AF nanocomposites in thin film form and their microstructures as a function of both film thickness and metal loading. The electrical conductivity of metal/polymer composites can be divided into different zones, a low-conductivity zone, followed by a transition zone during which the conductivity increases significantly with thickness and metal concentration, and a saturation zone, during which the conductivity levels off.

At thicknesses below about 150 Å, individual metallic nanoparticles separated by the polymeric matrix are observed. The electron conductivity in this dielectric zone is dominated by thermally activated electron tunneling from one metal island to the other [2]. Above the percolation threshold, continuous metallic pathways exist throughout the polymeric matrix. However, the conductivity of the composite is lower than that of pure metal films because the electron mean free path is greatly reduced due to the inclusion of the dielectric. In the percolation zone, adjacent metallic nanoparticles undergo extensive coalescence resulting in large irregular nanoclusters. The rapid increase in electrical conductivity is the result of an increased connectivity of the metallic nanostructures which is dependent on the film thickness and the metal loading. The electron conductivity in this transition zone can be described by the percolation model, with the dominant charge transfer mechanism changing from thermally activated electron tunneling to metallic conduction [2].

## 2. Experimental Details

Plain soda-lime glass microscope slides (75 x 25 x 1 mm$^3$) were purchased from Fisher Scientific and cut in half. Prime grade Si (100) wafers with a diameter of 2" were obtained from Silicon Quest International and cut into small squares of about 8 x 8 mm$^2$. These silicon substrates were used to characterize composite samples by scanning electron microscopy (SEM). Both types of substrates were sonicated in acetone and methanol solvents, rinsed with copious amounts of deionized water, and finally dried in a nitrogen stream. Silver wires of 99.999% purity (metal basis) were obtained from Alfa Aesar and used as is. An amorphous fluoropolymer, Teflon® AF 2400 (poly[4,5-difluoro-2,2-bis(trifluoromethyl)-1,3-dioxole-co-tetrafluoroethylene]), was acquired from DuPont and used as is. Teflon® AF has excellent optical transparency (>95%) in the visible and near-infrared wavelength ranges combined with good mechanical and chemical stability.

Ag/Teflon® AF nanocomposite films were fabricated by vapor-phase co-deposition in a high-vacuum chamber (base pressure of 10$^{-7}$ torr). A schematic representation of the experimental setup is depicted elsewhere [4]. The synthesis chamber is equipped with an electron-beam evaporator (Mantis Deposition Ltd.) which contains four individually controlled pockets, allowing for sequential or simultaneous thermal evaporation of up to four different materials. Since metals and polymers have weak interactions, the substrate was kept at an elevated temperature of 80°C to enhance the silver adsorption on Teflon® AF [17]. An Inficon quartz crystal microbalance (QCM) was utilized to estimate the film deposition rate and thickness. The Teflon® AF deposition rate was kept constant at 0.1 Å/s and the silver deposition rate was varied to adjust the metal loading. The QCM was calibrated with AFM measurements on a continuous thick film for each material. Specimens for electrical measurements were deposited onto glass substrates because of their smooth surface and dielectric nature. Two co-planar metal wire electrodes were placed on the substrate surface, 10 mm apart and glued to the substrate with water-based conductive graphite adhesives. A dielectric mask made from a Teflon® sheet with a 5 mm x 10 mm opening was used to define the deposition area. Samples for SEM characterization were deposited onto silicon substrates to reduce charging during the microstructure imaging. The silicon substrates are covered with a layer of amorphous native oxides (0.6-2.0 nm thick) [18], and thus have a surface similar to that of the glass slides which were used for the electrical characterization of the composite samples.

A Keithley 2400 source-meter unit (SMU) interfaced with a computer for automatic data collection was used for the electrical current measurements. These measurements were performed in-situ with the sample at the deposition temperature and under vacuum. A coaxial cable feedthrough was used to minimize external interferences on the measurements, especially at low current levels. The SMU has a current and voltage range of 10 pA to 1.05 A and 1 µV to 210 V, respectively.

The morphology of the film surfaces was examined using an FEI Sirion 200 field emission SEM (FESEM) operating at 15 keV. Image analysis software (ImageJ) was used to determine structural film parameters such as nanoparticle dimensions from the SEM images. Cross sectional samples for morphology and thickness information were prepared using a Reichert-Jung ultra microtome from Cambridge Instruments GmbH and analyzed with a JEOL JEM 1200 EX transmission electron microscope (TEM) operating at 100 keV.

## 3. Results and discussion

### 3.1 Synthesis and microstructure characterization of Ag/Teflon® AF nanocomposite films

Fig. 1 shows the chemical structure of Teflon® AF which is composed of tetrafluoroethylene (TFE) monomers and dioxole monomers. Teflon® AF 2400 consists of 87 mol% of dioxole and 13% of TFE [19]. Teflon® AF was selected as the polymeric matrix because of its outstanding optical clarity and high transmittance in the visible and infrared wavelengths regions, which is due to its amorphous nature. Ag/Teflon® AF nanocomposite films were synthesized using a vapor phase co-deposition process. The formation of metal clusters in a polymer matrix is a complex dynamic process. Upon thermal evaporation, Teflon® AF undergoes fragmentation, and re-polymerization occurs during its deposition on the substrate [20, 21]. Some of the vaporized metal atoms that condense on the polymer surface can migrate along the surface or into the polymer layer and form larger metal islands and clusters. Other metal atoms will desorb from the surface due to their weak interaction with the polymer. The ratio of the adsorbed metal atoms to the total metal atoms arriving at the substrate surface is given by the condensation coefficient. The condensation coefficient of silver for Teflon® AF is about 0.16 [22], implying a relatively weak interaction between silver and Teflon® AF. During the vapor phase co-deposition the metal concentration can be controlled by simply varying the deposition rate ratio of silver to Teflon® AF monitored by a QCM.

Figures 2, 3 and 4 illustrate a series of electron micrograph images of Ag/Teflon® AF nanocomposites with low to high metal loadings (50%, 73%, and 83%, respectively) for various film thicknesses. At thinner film stages below about 150 Å, the silver nanoparticles are well dispersed in the Teflon® AF matrix and isolated individual islands are the dominant features (Figs. 2a, 3a and 4a). The median particle size is about 5-9 nm depending on the metal concentration as shown in Fig. 5. As the nanocomposite film thickness increases, the size of the metal particles also increases. Due to their mutual affinity and weak interaction with the surrounding Teflon® AF matrix, newly arriving silver atoms and already embedded silver clusters or small particles tend to merge into larger metal particles through migration and diffusion to minimize the system energy during the dynamic vapor phase co-deposition process. This growth process can be further facilitated by the extra thermal and kinetic energy supplied by the incoming atoms during the vapor condensation [23]. The increase in particle size and concomitant decrease of interparticle spacing causes some neighboring metallic nanoparticles to coalesce, leading to the formation of metallic nanoclusters and the growth of fractal structures (Figs. 2b, 3b and 4b). A significant increase of the size of the metal nanocrystals is observed at this stage, with a median diameter of about 10-16 nm at 450 Å. The fractal growth also results in nanocrystals with more irregular shapes and broader size distributions as a result of mergers of adjacent metallic nanocrystals. Thereafter, although the films continue to build up, the nanocrystals grow very slowly in size (11-20 nm at a film thickness of about 1500 Å in Figs. 2c, 3c and 4c) indicating that a relatively stable microstructure has been reached.

The embedded silver nanoparticles are in a multilayer configuration as shown in the cross-sectional TEM images (Fig. 2d and the left image in Fig. 3b). The similar size of the nanocrystals throughout the film indicates that silver atoms have a pronounced diffusion in the Teflon® AF matrix. This is further confirmed by another sequential layer-by-layer deposition with Teflon® AF sandwiched between silver films, in which silver clusters were found throughout the entire Teflon® AF middle layer. The relatively long diffusion range probably originates from the weak interaction between silver and Teflon® AF as well as the amorphous nature of Teflon® AF, and may be further facilitated by the latent heat released during the vapor condensation.

As illustrated in Fig. 5, at similar film thicknesses, nanocomposites with higher metal content tend to have larger embedded silver nanoparticles and broader size distributions. For very thick films (above 4000 Å), particle aggregates with micrometer sizes were observed as shown in Fig. 3d. Close-up micrographs (inset in Fig. 3d) within these large aggregates indicate that they are composed of finely dispersed metallic nanocrystals in a polymorphous arrangement. These metallic nanoparticles do not coalesce into larger single crystals or grains due to the presence of the surrounding dielectric matrix. This matrix is very effective in preventing large grain formation during film growth. These superstructures are often observed in heterogeneous system because of the very dissimilar nature of the involved components. Since noble metals have cohesive energies at least two orders of magnitude higher than those of polymers [24] and their solubility is low in polymers, they have a strong tendency to aggregate due to their similarity. In addition, the polymers encasing the metal nanocrystals can further facilitate the formation of giant clusters [25].

### 3.2. Electrical conductivity of Ag/Teflon® AF nanocomposite films

Figure 6 depicts the normalized electrical current in Ag/Teflon® AF nanocomposite films as a function of film thickness under constant external potential (1V) for various metal loadings. For nanocomposites with metal loadings between 30% and 80%, the conductivity can be divided into three distinct zones indicated with gray dashed lines in Fig. 6. In the first zone ($C_I$ in Fig. 6) with nanocomposite film thicknesses of less than 150 Å, the samples exhibit dielectric behavior with a measured current of about 50-100 pA. This low current is due to the isolated nature of the silver nanoparticles with small sizes and relative large separations in the dielectric Teflon® AF matrix (Figs. 2a, 3a and 4a). As the films grow thicker, the measured current increases rapidly. This increase is due to the interconnectivity and the formation of a fractal metallic structure as shown in Fig. 2b, 2c and 2d. This rapid increase in the current in the transition zone ($C_{IIA}$ in Fig. 6) also corresponds to the apparent jump in the size of the nanocrystals in Fig. 5, indicating the inherent dependence of the electrical conductivity on the composite microstructures. The slope, an indicator of the rate at which the current changes, becomes slightly steeper as the metal content increases. However, higher metal content extends the transition zone before reaching the plateau, which assures the films with higher metal content to reach higher current levels. At film thicknesses of about 375-450 Å the rate at which the current increases slows down into a relatively flat process (Zone $C_{IIB}$ in Fig. 6). Electron microscopy investigation shows that the size of the nanocrystals also reaches a plateau, further confirming the inherent correlation between film microstructures and conductivity.

For metal loadings below about 30%, the conductivity changes slowly and smoothly while the films continue to grow thicker. No apparent transition zone of a rapid current change is observed. At these low metal contents, the separation between silver nanoparticles by the polymer matrix is so large as to be out of their range of diffusion and migration since they are relatively immobile in contrast to atomic silver [26, 27]. The chance to become contacted or coalesced to form an interconnected metallic pathway is very low due to the effective hindrance of the surrounding dielectric polymers.

At metal concentrations of 90% and larger, a second transition (Zone $C_{IIC}$ in Fig. 6) is observed, leading the nanocomposite conductivity to reach the flat metallic continuum regime ($C_{III}$ in Fig. 6). At present, it is believed that this second threshold originates from the formation of extended metallic pathways, as the large percentage of metallic surfaces without polymer encasings leads to more direct metal-to-metal contacts. Continuous conductive percolation paths form at this stage. Accordingly, the dominant charge transport mechanism switches from tunneling to metallic conduction, which is evidenced by the comparable current level to pure silver films (green plot in Fig. 6). However, silver in composites still keeps the morphology of small nanocrystals (Fig. S1 of the supplementary information), which is attributed to the strong anti-coarsening effect of the encasing polymer.

The electrical properties of Ag/Teflon® AF nanocomposites are closely linked to the morphology of the metallic nanoparticles. A conduction model for island metal films, in which thermally activated tunneling is the dominant conduction mechanism, can be applied to polymer-metal composites below the percolation threshold [28, 29]. As the films grow thicker, initially isolated particle islands start to coalesce, resulting in a significant decrease of the interparticle spacing and the formation of fragmented nanoclusters on a nanometer to micrometer scale (Fig. 2b, 3b and 4b). The decreased interparticle spacing can significantly increase the electron tunneling probability, and fragmented metallic paths with more efficient Ohmic conduction exist within nanoclusters. In other words, some portion of tunneling is replaced by the more efficient Ohmic conduction in the nanocomposite films. All foregoing factors initiate the giant increase of electrical conductivity in first transition zone ($C_{IIA}$ in Fig. 6), which is a direct consequence of the nanostructure evolvement. At film thicknesses above 375-450 Å, the interconnection between neighboring nanoparticles reaches a relatively stable level and the conductivity changes little with the increase of film thickness. Although the conductivity enters into a plateau (Zone $C_{IIB}$ in Fig. 6), the dominant mechanism is still thermally activated tunneling since no continuous path has been formed and the current is kept on the level of nA to µA depending on metal concentration. Complete metallic continuum (Zone $C_{III}$ in Fig. 6) is only observed at higher silver loadings above about 90% and film thicknesses larger than about 1350 Å, following a second rapid increase in current (Zone $C_{IIC}$ in Fig. 6) as a result of the formation of percolation path. Because the metal loadings are very high, the chance for nanoparticles to directly contact with each other becomes so frequent that a complete metallic network forms with dielectric polymer inclusions. Accordingly, the dominant charge transport changes from tunneling to Ohmic conduction.

Fig. 6 also compares the normalized current vs. thickness plots for pure silver films and silver-doped Teflon® AF nanocomposite films. There are several distinct differences. In the first zone of isolated nanoparticles (Zone $S_I$ of Ag and Zone $C_I$ of composites in Fig. 6), both have similar currents on the order of a few pA. However, the first upturn transition point occurs at significantly thinner films for pure silver (75-100 Å) than for Ag/Teflon® AF nanocomposite films (150-200 Å). The interparticle space between silver nanoparticles contains vacuum and the particles grow in a Volmer-Weber mode (faster growth in lateral direction than in normal direction) [30], which quickly increases the particle size and decreases the interparticle spacing. For composite films, the interparticle space is filled with Teflon® AF. The encasing polymer limits the coalescence of the silver nanoparticles and prevents direct contact between them. In this region (Zone $S_{II}$ of Ag and Zone $C_{IIA}$ of composites in Fig. 6), the increase of current with increasing thickness occurs much faster for silver films than for Ag/Teflon® AF nanocomposite films. During this transition process, the silver films form a fully percolated metallic network and the conduction mechanism changes from tunneling to Ohm's law. For Ag/Teflon® AF nanocomposites however, tunneling is prevalent throughout the first transition stage ($C_{IIA}$ in Fig. 6). In the subsequent stage (plateau regions of Zone $S_{III}$ of Ag and Zone $C_{IIB}$ of composites in Fig. 6), silver films form a metallic continuum. For composite films, tunneling is still dominating the charge transfer where the current is at the level of nA to µA for low or moderate metal loadings (30-80%). No metallic continuum forms at this point for nanocomposite films. Only at very higher metal loadings (>90%), does a second transition (percolation of Zone $C_{IIC}$ in Fig. 6) occur, leading to the fast current increase to a level ($C_{III}$ in Fig. 6) comparable to that of pure silver film ($S_{III}$ in Fig. 6). At this point the embedded silver nanoparticles complete the percolation and eventually form a metallic continuum with dielectric polymer inclusions. However, the conductivity of the composites is lower than that of pure metals because the reduced electron mean free path induces additional electron scattering. It can be deduced that the polymer matrix is a very important factor in silver nanocrystals growth, keeping them from merging into coarse grains. At a fixed silver loading, the polymers significantly stretch the percolation period for nanocomposites (Zones $C_{IIA}$, $C_{IIB}$ and $C_{IIC}$) with respect to pure silver films (Zone $S_{II}$) in terms of film thickness, and only films with very high metal loadings eventually form a fully interconnected metallic network. The film conductivity is closely related to the morphology which is a function of both film thickness and metal concentration.

The normalized electrical current as a function of silver loadings at various film thicknesses under 1V bias is shown in Fig. 7, (100% silver loading refers to pure silver films of the same thickness as the nanocomposites). The connection of representative data points with dashed lines implies that they assume (part of) the typical description of conductivity vs. concentration based on the percolation model as shown for the silver plot in Fig. 6. According to our previous electrical measurements on pure silver films [31], the percolation threshold in silver films occurs at thicknesses larger than 160-180 Å (Fig. 6). Below this thickness, a continuous path can not be formed in nanocomposites. Beyond this thickness, percolation occurs as the silver content continues to increase and the electrical conductivity finally enters into the metallic regime. The percolation zone slightly shifts toward lower metal contents as the composite films grow thicker. This trend appears to continue as the system size increases [32, 33].

A representative illustration is provided in Fig. 8 to investigate how well Ohm's law is obeyed for the nanocomposite films with discontinuous metallic components. The voltage potential was swept from 0~2 V with a 0.1 V increment. Higher potentials were not adopted in order to avoid excess stress on the films. A decreased conductance was observed for nanocomposite films with isolated particle islands, which was also observed in pure silver films with similar morphology [31]. The charging effect of far separated islands may partly contribute to this phenomenon. Linear behaviors were observed (Fig. 8b, c), at least at low voltages, for nanocomposites which showed a high degree of coalescence. However, the dominant transport mechanism is still governed by electron tunneling since the metallic continuum has yet to be formed with very high metal loadings.

## 4. Conclusion

The electrical properties of Ag/Teflon® AF thin film nanocomposites were studied under an applied external constant potential as a function of film thickness at various metal loadings. The electrical properties are directly dependent on the morphology of the silver nanoparticles embedded in the Teflon® AF matrix. Composite films with a very low metal concentration and metallic nanoparticles well dispersed within the polymer matrix, show dielectric characteristics. Initially isolated nanoparticles in composite films with a moderate metal loading are able to coalesce into fractal nanoclusters as the films grow thicker, leading to a giant increase of conductivity before it levels off. No continuous electrical pathway has formed at this stage, and thermally activated tunneling dominates the charge transport. At very high metal concentration, metallic nanoclusters are able to initiate and accomplish the percolation process and eventually form the metallic continuum. This continuous metallic pathway leads to a second rapid increase in conductivity and the switch to Ohmic electron transport. The resulting conductivity is lower than that of pure metallic films due to the presence of dielectric polymer inclusions which significantly reduce the electron mean free path.

## ACKNOWLEDGMENT

This work was supported by ARO Grant W911NF-06-1-0295 and by ONR Grant N00014-03-1-0247.

**Figures**

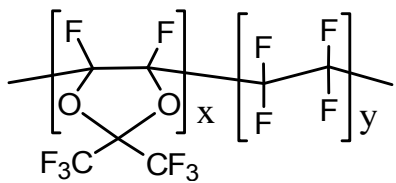

Fig. 1. Chemical structure of Teflon® AF, composed of tetrafluoroethylene monomers and dioxole monomers in a x:y ratio (see text for details).

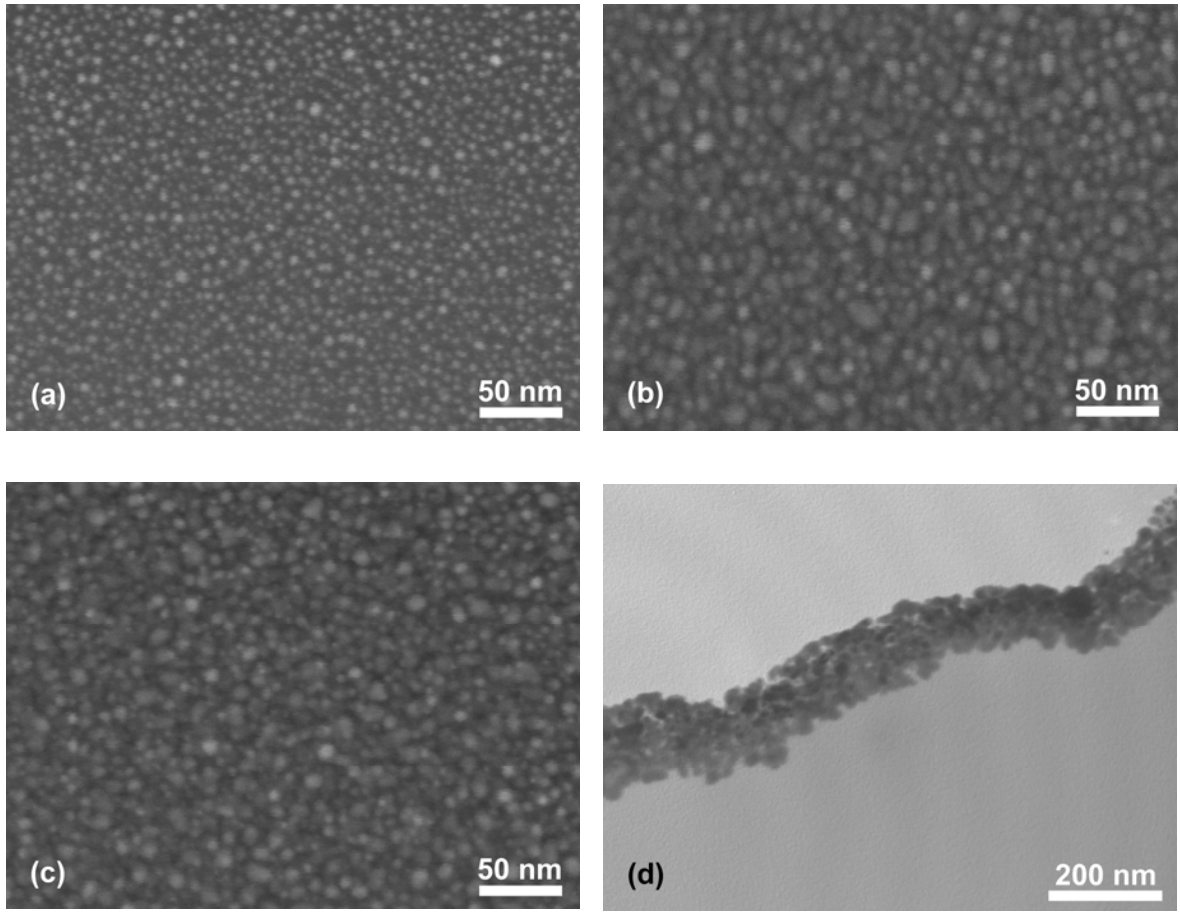

Fig. 2. Typical microstructural evolvement of Ag/Teflon® AF nanocomposites with a Ag loading of about 50% as a function of film thickness. (a) ~150 Å, (b) ~410 Å, (c) ~1500 Å, (d) ~1500 Å.

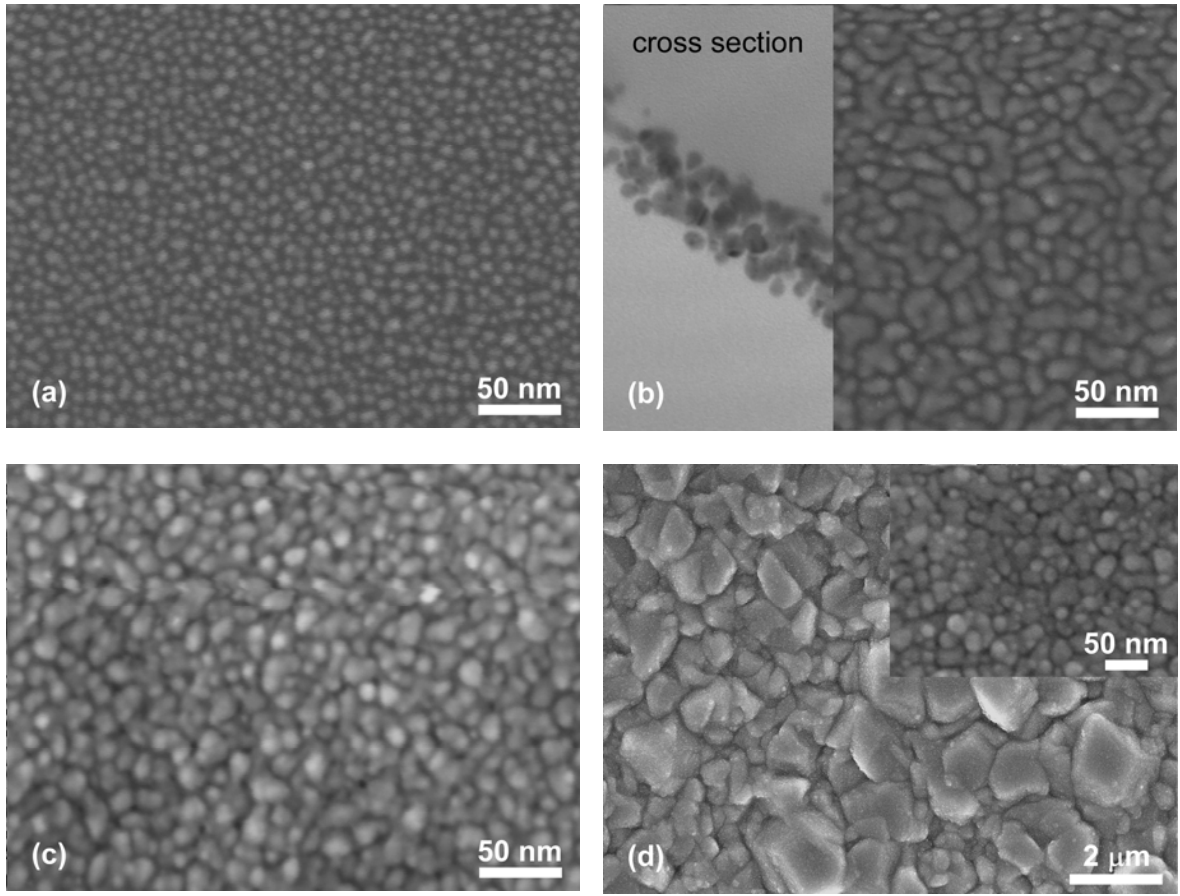

Fig. 3. Typical microstructural evolvement of Ag/Teflon® AF nanocomposites with a Ag loading of about 73% as a function of film thickness. (a) ~150 Å, (b) ~450 Å, (c) ~1710 Å, (d) ~6650 Å.

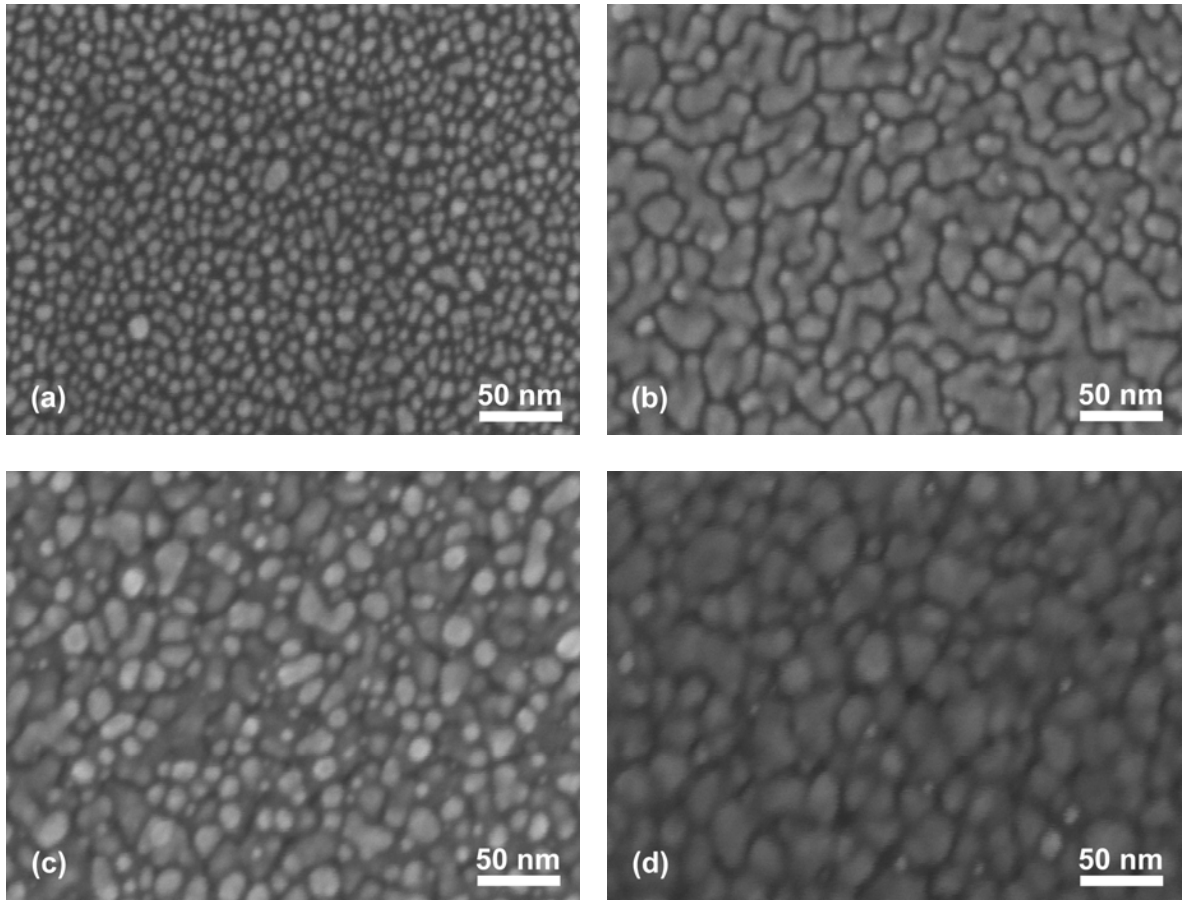

Fig. 4 Typical microstructural evolvement of Ag/Teflon® AF nanocomposites with a Ag loading of about 83% (a, b, and c) and 95% (d) as a function of film thickness. (a) ~150 Å, (b) ~450 Å, (c) ~1530 Å, (d) ~3000 Å.

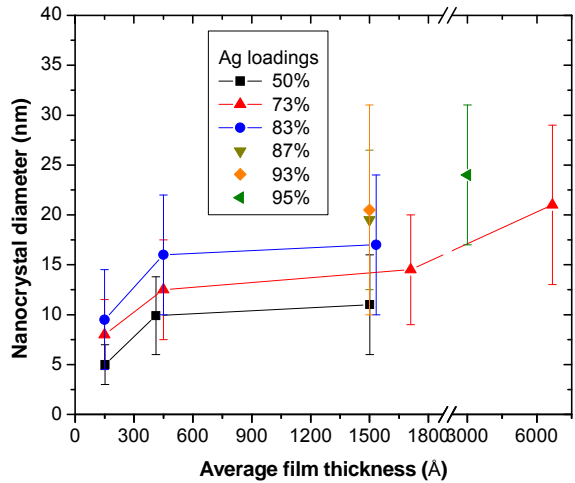 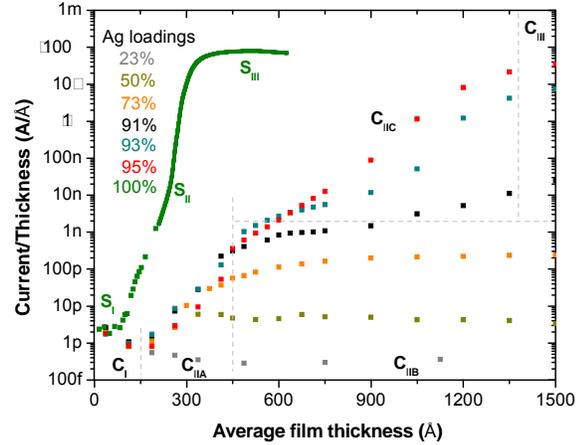

Fig. 5. Average nanocrystal diameter of Ag nanoparticles in Teflon® AF matrix. Upon coalescence, relative large crystal size increases were observed. Thereafter, the nanocrystal size increased only slightly as the composite films continued to build up.

Fig. 6. Normalized electrical conductance of Ag/Teflon® AF nanocomposite films as a function of film thickness with various metal loadings under 1V external potential. The silver plot is included for comparison under the same measurement condition.

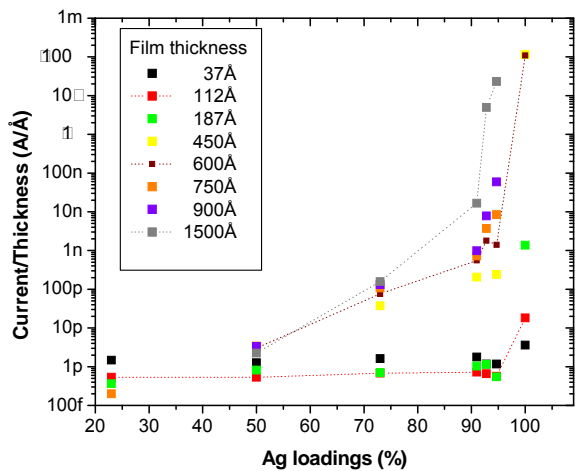
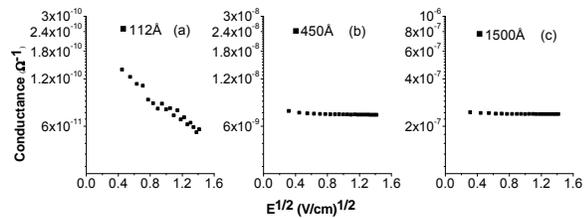

Fig. 7. Normalized electrical conductance of Ag/Teflon® AF nanocomposites as a function of Ag concentration at various film thicknesses. The current was measured under 1V external potential.

Fig. 8. Log conductance vs. the square root of applied external field at different composite film thickness. Ag concentration is ca. 73%.